\documentclass[a4paper]{article}
\usepackage{graphicx}
\usepackage{amssymb,amsmath,bm}
\usepackage{textcomp}

\title{Individual identity in songbirds: signal representations and metric learning for locating the information in complex corvid calls}

\makeatletter
\def\name#1{\gdef\@name{#1\\}}
\makeatother \name{{\em Dan Stowell$^1$, Veronica Morfi$^1$, Lisa F. Gill$^2$}}

\author{Dan Stowell$^1$, Veronica Morfi$^1$, Lisa F. Gill$^2$ \\ \\
$^1$Machine Listening Lab, Centre for Digital Music, \\
Queen Mary University of London, UK \\
  $^2$Max Planck Institute for Ornithology, Seewiesen, Germany \\
{\small \tt \{dan.stowell, g.v.morfi\}@qmul.ac.uk, lgill@orn.mpg.de}
}

\begin{document}

  \maketitle
  \begin{abstract}%
Bird calls range from simple tones to rich dynamic multi-harmonic structures. The more complex calls are very poorly understood at present, such as those of the scientifically important corvid family (jackdaws, crows, ravens, etc.). Individual birds can recognise familiar individuals from calls, but where in the signal is this identity encoded? We studied the question by applying a combination of feature representations to a dataset of jackdaw calls, including linear predictive coding (LPC) and adaptive discrete Fourier transform (aDFT). We demonstrate through a classification paradigm that we can strongly outperform a standard spectrogram representation for identifying individuals, and we apply metric learning to determine which time-frequency regions contribute most strongly to robust individual identification. Computational methods can help to direct our search for understanding of these complex biological signals.   \end{abstract}

  \section{Introduction} 

Bird vocalisations are highly complex. They are often analysed as sinusoidal or harmonic sounds, or as spectrotemporal ``patches'' \cite{Stowell:2010e}, but in general this can obscure their rich structure: unlike humans, songbirds have two sets of vocal folds, which they can use simultaneously; and they also have muscles specialised for rapid pitch modulation \cite{Riede:2010}. Songbird species make use of these abilities in different ways, not all of which are fully understood.
These complexities pose problems for signal processing paradigms borrowed from the study of speech or music.
Yet bird vocalisations are an important area of scientific study: from behavioural studies we know that they can contain information about species, about individual identity, and more, and so we seek signal processing methods that can represent that information in a form suitable for analysis.
Corvids, a family of songbirds that includes ravens, crows, jays and other species, have been the focus of much research, due to their social complexity and remarkable cognitive skills. But unlike in other songbirds, their vocalisations have not been as extensively studied, maybe because they mainly produce short, non-tonal vocalisations (calls) that are often structurally complex and may involve the two-voice phenomenon. However, their vocal complexity, in combination with high levels of sociality and cognition has made corvids a suitable target for studying vocal recognition \cite{Kondo:2010}.
Jackdaws {\em(C. monedula)} are highly vocal, group-living corvids that breed in colonies and form strong, lifelong pair bonds. They are highly vocal and use a variety of vocalisations to maintain contact and communicate with their conspecifics \cite{Cramp:1994}. Recently, it has been suggested that pair members are able to recognise each other's contact calls \cite{Gill:2015ibac}.
But are we able to discriminate individuals by analysing these challenging vocal signals?
And which parts of the signals carry the individual information?
To explore these questions, we analysed a dataset of calls coming from 20 individually recorded jackdaws,
using alternative signal representations together with automatic classification and metric learning.

\begin{figure}[tp]
	\centering
	\includegraphics[height=6.3cm,clip,trim=0mm 69.5mm 130mm 6mm]{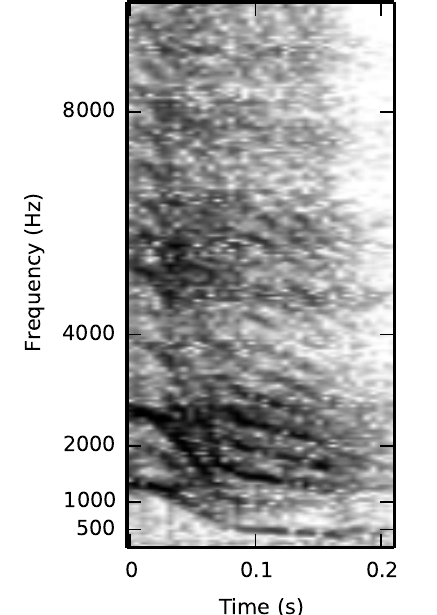} %
	\includegraphics[height=6.3cm,clip,trim=12.2mm 69.5mm 130mm 6mm]{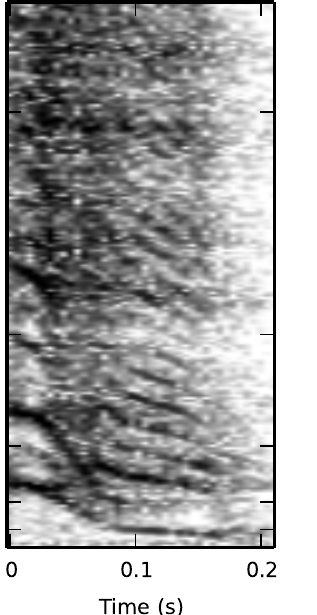} %
	\caption{Spectrogram of a single jackdaw call (left) and of its LPC residual (right).}
	\label{fig:specgrams}
\end{figure}

  \section{Method} 

\subsection{Data collection}

The dataset came from 20 adult, hand-raised, group-housed jackdaws (in accordance with the 2010/63/EU European directives for the protection of animals used for scientific purposes). Individual vocalisations were recorded using a microphone (TC20, Earthworks, USA) and solid-state recorder (PMD661, Marantz, Japan, at a sampling rate of 48000 Hz) while the birds were temporarily held in social isolation (for max. 32 minutes) in a sound chamber (2.0 x 2.0 x 1.3 m, fitted with white acoustic foam), as part of a different study. Contact calls were classified and clean recordings selected by visual and acoustic assessment. Here, we analysed a total of 1156 calls (3 to 93 per individual, median = 60), with a median duration of 0.272 seconds (standard deviation: 0.047, min: 0.160, max: 0.510 seconds). 
An example call is shown in Figure \ref{fig:specgrams} (left).

\subsection{Signal processing}

Our aim was to produce spectrogram representations that allowed individual identity information to surface most readily.
For standard spectrograms we used a frame size of 1024 with Hann windowing and 75\% overlap.

As a potential improvement on this baseline we evaluated linear prediction (LPC) analysis,
widely used in speech processing as a step which aims to separate the effects of the glottal source (for songbirds, the \textit{syringeal} sources) from the vocal tract filter \cite{Makhoul:1975lpc}.
To each call we applied LPC analysis of order 10 to the whole audio clip.
Songbirds might encode individuality in fixed/gestural aspects of the syrinx oscillation and/or the vocal tract setting.
We therefore used the LPC residual (which is an approximation to the syringeal source signal) in the same way as we would use the original raw audio, for spectrogram processing (see Figure \ref{fig:specgrams} right),
and we also used the LPC spectrum (an approximation to the vocal tract filter) separately as an alternative spectral representation of each call.

Separately we explored the use of the adaptive Discrete Fourier Transform (aDFT) proposed in \cite{Morfi:15}.
This was originally introduced as an alternative, adaptive method to compute a spectrogram and the sinusoidal parameters used by the ``adaptive harmonic model'' (aHM) in human speech analysis and synthesis.
The aDFT is similar to the Discrete Fourier Transform (DFT) but uses a frequency basis that can completely follow the variations of the fundamental frequency $F_0$ throughout a recording, in contrast to the constant frequency basis of DFT.
This time-varying $F_0$ is estimated as part of the algorithm,
which thus offers the potential of a more accurate time-frequency representation. The harmonics can be easily traced and are typically more prominent even in mid/high frequencies, compared to those in the DFT spectrogram.

In order to apply the aDFT to our dataset only one modification of the original algorithm was needed.
The $F_0$ variations for human speech are considered to be between 40 and 700 Hz,
but songbirds have an even higher and wider range.
Hence, an $F_0$ range between 80 and 2000 Hz was used for the estimation of the $F_0$ variations throughout each recording.
The values of the $F_0$ curve that was used as the frequency basis for aDFT were then divided by 2, because of the potential importance of sub- and inter-harmonics.
Furthermore, we tested this aDFT with and without the adaptive iterative refinement algorithm, also proposed in \cite{Morfi:15},
which iteratively refines the $F_0$ estimate used to produce the aDFT spectrogram. 
We therefore tested `unrefined' and `refined' versions.

\begin{figure}[tp]
	\centering
	\includegraphics[height=6.3cm,clip,trim=0mm 69.5mm 130mm 6mm]{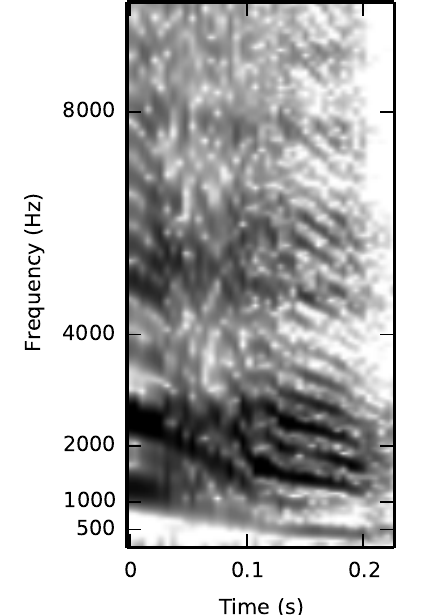} %
	\includegraphics[height=6.3cm,clip,trim=12.2mm 69.5mm 130mm 6mm]{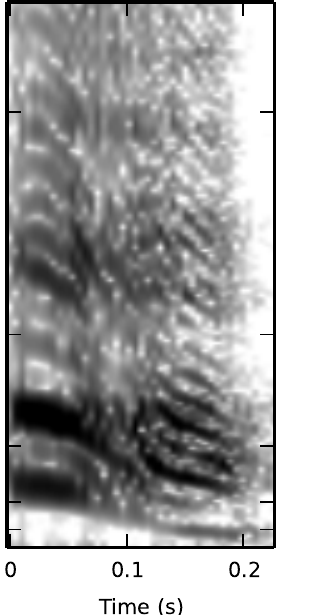} %
	\caption{aDFT spectrogram of a single call: using unrefined (left) and refined (right) $F_0$ estimate.}
	\label{fig:specgramsadft}
\end{figure}

The aDFT is designed to improve the characterisation of frequency-modulated harmonic sounds,
and so is a candidate for analysis of songbird vocalisations
(although it does not directly model two-voice phenomena).
The aDFT produces spectrograms with a varying rather than regular temporal frame rate. In order to perform pairwise comparisons between spectrograms, and to maximise comparability against the standard spectrogram representation, we resampled the aDFT spectrograms onto a regular grid (by nearest-neighbour interpolation) at the same frequency and time resolution as the standard spectrogram (Figure 	\ref{fig:specgramsadft}).

\subsection{Classification tests}

We used a classification paradigm to evaluate the various signal representations.
Our aim was not to achieve the best classification possible,
but to probe which representations gave the clearest indications of individual identity.
For this reason, we did not use an arbitrarily powerful classifier,
but the well-known and simple $k$-nearest-neighbours (kNN) classifier (with $k=3$).
Data that work well with a kNN classifier will have their classes in compact, well-separated clusters.
However, the kNN classifier makes few assumptions about class distributions beyond that: it is tolerant of multimodal classes, and of classes having different variances, as long as they remain compact and well-separated.
kNN also naturally encompasses multi-class data and not just binary-labelled data.
Since we were working with multi-class data in a high-dimensional space (spectrogram pixels)
about which relatively little is known (e.g. class multimodality),
kNN was preferred over other simple discriminators such as linear discriminant analysis.

The classifier was applied to pairwise distances between spectrograms,
evaluated using four different distance metrics: Euclidean and Manhattan, each applied to magnitudes and to log-magnitudes.
The sound files were all aligned by their onset.
In order to reduce any problems due to relative misalignment
when measuring pairwise distances, we allowed a relative movement of $\pm 20$ ms and used the alignment that gave the minimum distance.

We therefore obtained classification results using data preprocessed in various ways:
with/without LPC analysis,
with a standard spectrogram or aDFT (and aDFT could be with/without refinement of its $F_0$ estimate),
with/without log-transformed magnitudes,
with Euclidean or Manhattan distance.
We evaluated the contributions of these factors using a mixed-effects statistical model (GLMM, with call recording ID as the random factor), to determine whether each made a significant difference to performance, and to check for interactions between the factors.
In order to help understand the effects of the various signal-processing choices, we also visualised the pairwise distances using t-SNE \cite{VanderMaaten:2008}.

The above investigations help to ensure that we choose signal processing methods that elucidate individual-identifying information if it is present,
and through LPC help us to consider one facet of the question of `where' the information lies.
It is natural to ask to what extent information lies in the onset or the decay of the call, in the fundamental or the harmonics---the latter question being particularly pertinent when a call can simultaneously contain energy from the two syringeal sources.
Note that the kNN classifier depends fundamentally on the distance metric used, and in particular that the relative weighting of features can make a dramatic difference to results.
In the first test we have described, we varied the metric but did not vary the feature weightings, in effect telling the classifier to pay equal attention to every pixel.
We therefore complemented this approach by using the large margin nearest neighbours (LMNN) framework (a type of \textit{metric learning}) to derive an analysis of the relative importance of each spectrogram pixel.
The LMNN process learns a linear transformation of the data space in such a way as to maximise the separation of classes, specifically to optimise kNN performance \cite{Weinberger:2005}.
It can therefore improve on the basic kNN classification results;
however our focus was not on the improved classification score but on the linear projection that was learnt.
After LMNN was trained (using the Python \textit{metric\_learn} package)
we mapped the projection matrix back onto the spectrogram pixels, giving an overall importance weight for each pixel.
(Note that for implementation reasons we could not include the $\pm 20$ ms variable-alignment in the LMNN test.)
To visualise these results, we applied the ranking transformation to the importance weights (to normalise their dynamic range) and then plotted them in the same format as spectrograms.

  \section{Results} 

\begin{figure*}[t]
	\centering
-	\includegraphics[page=25,width=0.45\linewidth,clip,trim=0mm 0mm 0mm 0mm]{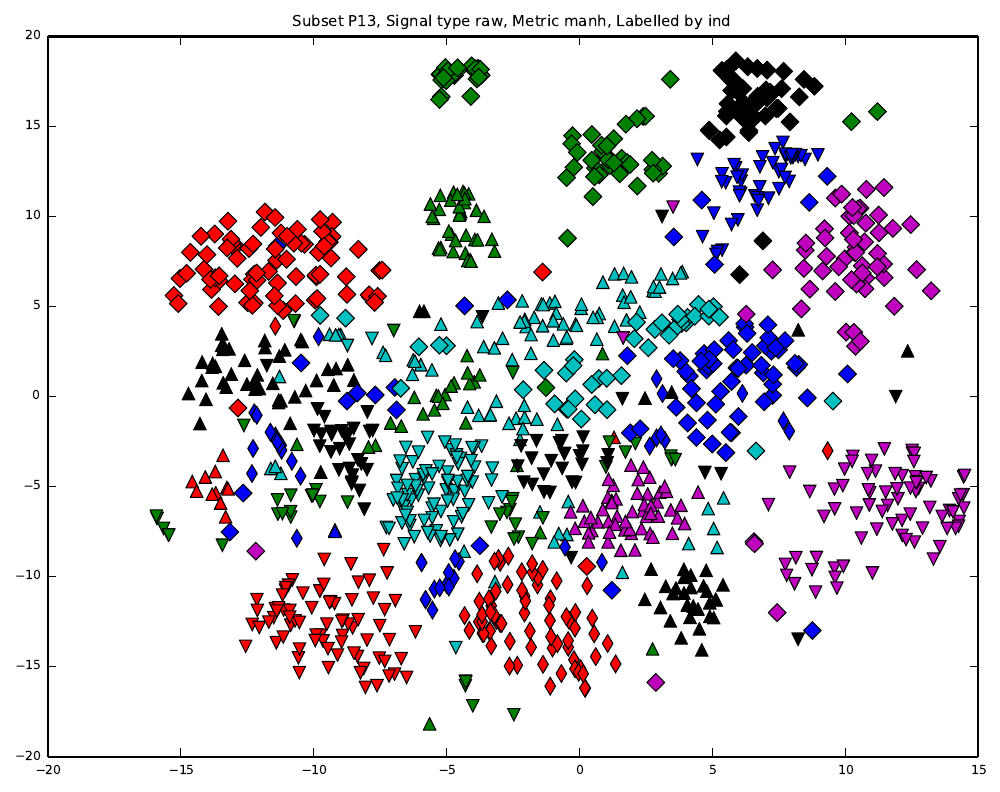}
	\includegraphics[page=13,width=0.45\linewidth,clip,trim=0mm 0mm 0mm 0mm]{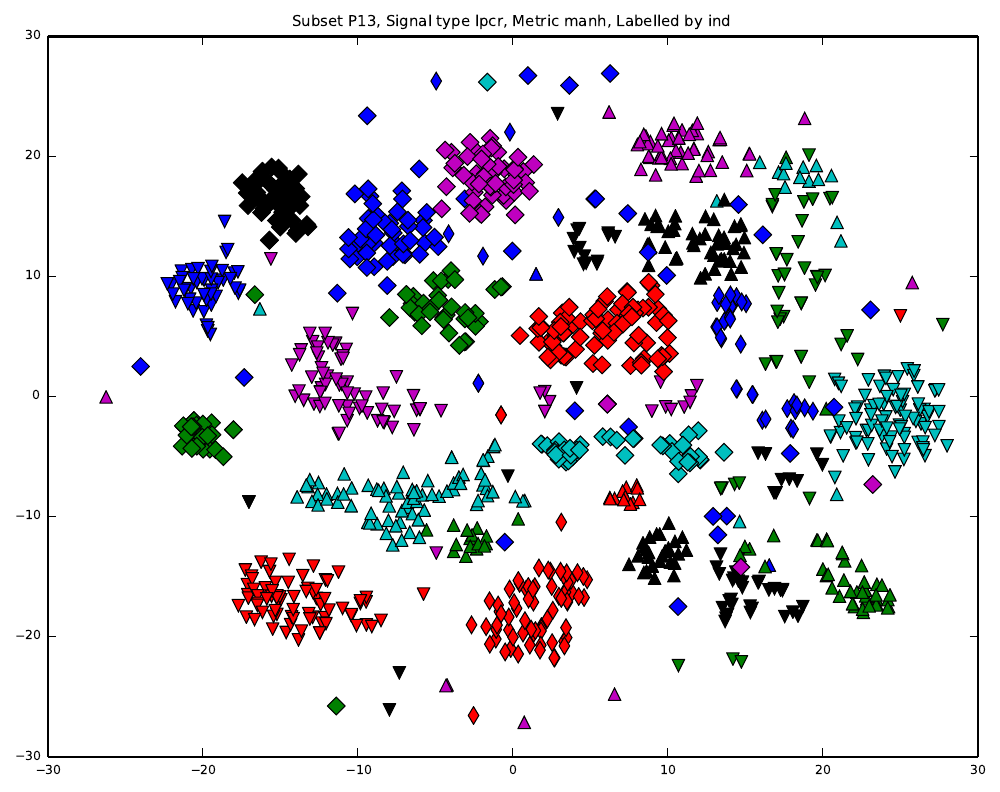}
	\caption{t-SNE distance plots for jackdaw calls under two different signal representations, in this case using Manhattan distance. Left: raw audio; right: LPC residual. Each marker type/colour represents a different individual. The axes have no direct interpretation.}
	\label{fig:distanceplots_P13_ind}
\end{figure*}

\begin{figure*}[tp]
	\centering
	\includegraphics[width=\linewidth,clip,trim=0mm 0mm 0mm 0mm]{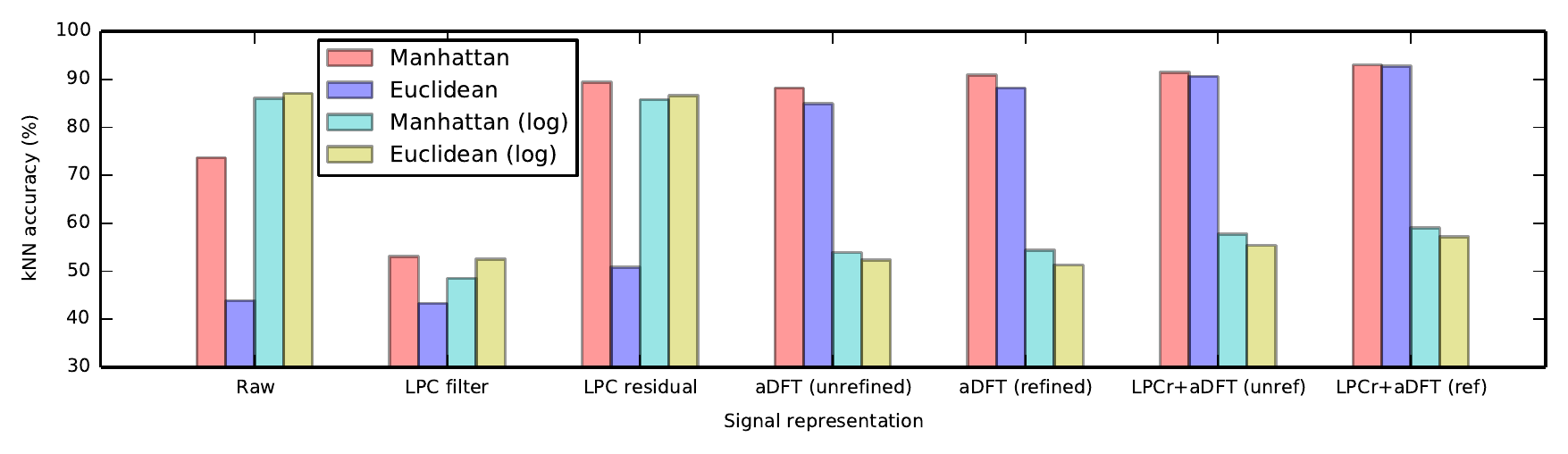}
	\caption{kNN classification results. 
    Chance level is 8.0\%.
    For each signal representation, results are shown for each of the four different distance measures.}
	\label{fig:classifresults}
\end{figure*}

Classification results show clearly that a dramatic improvement over a standard spectrogram representation is possible (Figure \ref{fig:classifresults}).
Our GLMM analysis found a significant effect of all our signal-processing interventions, but also significant interactions between all of them (as can be seen in the rather bimodal results of Figure \ref{fig:classifresults}),
so here we will not focus on the GLMM results in detail.
In general the Manhattan metric gave best results.
When using Manhattan distance, we found that the LPC residual \textit{and/or} aDFT led to a strong improvement from around 74\% to 90\% in individual identification;
yet the joint application of LPC and aDFT did not strongly improve results beyond that.
The aDFT spectrograms showed strong performance with either Manhattan or Euclidean distance.
The LPC \textit{filter estimate}, however, did not lead to strong classification, performing noticeably worse than the standard spectrogram.

The t-SNE plots give visual indications of the characteristics of the high-dimensional spaces produced by the variant analyses.
Figure \ref{fig:distanceplots_P13_ind} shows the strongest-performing metric (Manhattan distance) for the raw audio and the LPC residual; the LPC residual can be seen to enhance the per-individual clustering structure.
(Plots for aDFT were qualitatively similar to that for the LPC residual.)
Some of the classes appear multimodal, which may for example reflect an individual's use of differing call types.

\begin{figure}[tp]
	\centering
	\includegraphics[width=0.99\linewidth,clip,trim=0mm 0mm 0mm 0mm]{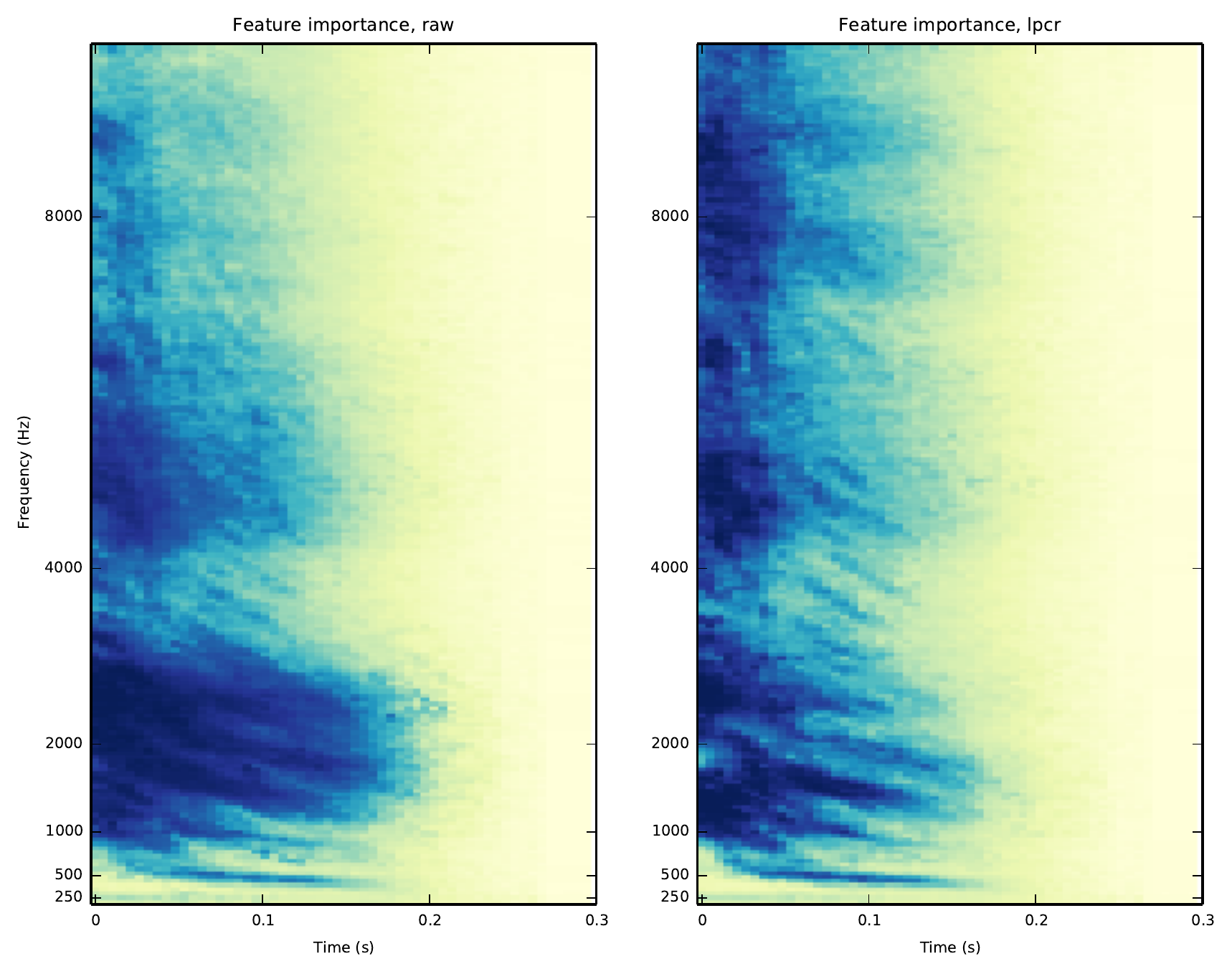}
	\caption{Feature importance maps derived from LMNN, for raw audio (left) or LPC residual (right). Darker pixels are higher-ranked in LMNN weighting.}
	\label{fig:hotpixels}
\end{figure}

Feature importance weightings derived from LMNN help to understand the relative importance of specific time-frequency regions, and they also offer clues as to what sort of benefit the LPC preprocessing gives (Figure \ref{fig:hotpixels}).
For both raw audio and LPC residual, the strongest-weighted pixels are concentrated soon after call onset, and place heavy emphasis on regions containing the fundamental (which in many cases includes a downward sweep from approx.\ 1 kHz) and its first one or two harmonics -- more broadly speaking, the region 1--3 kHz.
Features which attain more importance after LPC transformation are the higher frequencies near the onset.
LPC analysis inherently produces a residual with a whiter spectrum, and so for these sounds amplifies the upper harmonics (Figure \ref{fig:specgrams}). However this does not inherently lead the LMNN analysis to apply more weight to them; if the higher harmonics did not contain useful information we would expect LMNN to suppress them.

  \section{Discussion} 

The results here do not directly tell us about the production or perception processes involved in jackdaw contact calls, but they do give us information which helps to guide investigations into those processes, and also which helps us to design systems to extract information automatically from such vocalisations.
We have demonstrated that signal processing interventions can dramatically improve the automatic identification of individuals with a low-complexity classification algorithm,
which tells us that they can transform the signal into a format in which individuals' calls have more stable and repeatable character.

Our feature importance analysis suggests that the identifying information is concentrated soon after onset,
and spreads across the fundamental(s) and harmonics.
However, we note that for this particular analysis we could not vary the relative alignment of the signals, meaning that discriminative information in the tail-end of calls may not be apparent since calls are variable in length.

We found that both LPC and aDFT lead to representations that facilitate classification.
These are two rather different interventions, yet the improvements are not strongly additive: results when using both are only a little stronger than when using either.
LPC is a well-known technique, and relatively efficient to apply,
whereas aDFT and related representations are not very well studied,
and take much more computation than a standard spectrogram.
The cost-benefit ratio therefore speaks in favour of the LPC residual for the moment,
though we have shown that advanced adaptive spectrograms are worth exploring for their surfacing of information present in the signal.

The LPC residual, rather than the LPC filter,
showed the strongest connection with individual identity in our tests.
Linear prediction is often used for source-filter analysis,
with the residual interpreted as the glottal or syringeal source signal.
However, the LPC residual is also likely to normalise away any differences in the channel between bird and microphone, and so one should be cautious before interpreting the LPC filter component as purely representing ``the vocal tract''.
It also means that the improved recognition using the LPC residual could be interpreted as a normalisation rather than a decomposition.
However---contrary to that line of thought---in the present dataset recordings are controlled and the channel effects are stable,
although factors such as the orientation of a bird's head with respect to the microphone can still affect the received signal.
Nevertheless, our results suggest that the signal component from the syringeal source contains sufficient information for recognising individuals in this species,
and that this information is not just in the fundamental frequencies but also involves the overall harmonic structure of the source signal---%
a suggestion which should be compared against physiological and perceptual evidence in future.

  \section{Acknowledgements}

This work was supported by
EPSRC Early Career research fellowship EP/L020505/1.
LFG was funded by the Max Planck Society.
We also thank Auguste von Bayern for discussions and support during the experiments.

  \newpage
  \bibliographystyle{IEEEtran}

\bibliography{jackdawcallreps}

\end{document}